\documentclass[iicol,pdflatex,sn-mathphys-num]{sn-jnl}

\usepackage{graphicx}%
\usepackage{multirow}%
\usepackage{amsmath,amssymb,amsfonts}%
\usepackage{amsthm}%
\usepackage{mathrsfs}%
\usepackage[title]{appendix}%
\usepackage{xcolor}%
\usepackage{textcomp}%
\usepackage{manyfoot}%
\usepackage{booktabs}%
\usepackage{algorithm}%
\usepackage{algorithmicx}%
\usepackage{algpseudocode}%
\usepackage{listings}%
\usepackage[normalem]{ulem}
\usepackage{dblfloatfix}

\begin{document}

\title[Article Title]{Reprogrammable origami through bistable buckled hinges}


\author[1]{\fnm{Leon M.} \sur{Kamp}}\email{lkamp@seas.harvard.edu}

\author[1]{\fnm{Lucy} \sur{Liu}}

\author[1]{\fnm{Ella} \sur{McRitchie}}

\author[2]{\fnm{Damien} \sur{Rouchouse}}

\author[2]{\fnm{Orel} \sur{Mazor}}

\author[3]{\fnm{Davood} \sur{Farhadi}}

\author[1]{\fnm{L.} \sur{Mahadevan}}

\author*[1]{\fnm{Katia} \sur{Bertoldi}}\email{bertoldi@seas.harvard.edu}

\affil*[1]{\orgdiv{J.A.Paulson School of Engineering and Applied Sciences}, \orgname{Harvard University} \orgaddress{\city{Cambridge}, \state{MA}, \postcode{02138} \country{USA}}}

\affil[2]{\orgname{PSL University}, \orgaddress{\city{Paris}, \country{France}}}

\affil[3]{%
\orgdiv{Department of Precision and Microsystems Engineering},
\orgname{Delft University of Technology},
\orgaddress{
\street{Mekelweg 2},
\city{Delft},
\postcode{2628 CD},
\country{The Netherlands}
}}


\abstract{Origami structures typically have a multiplicity of folded states that are connected to a flat sheet, making the folding protocol for specific end shapes challenging to design and deploy. Here, we introduce reprogrammable origami hinges that use bistable buckled shims to reversibly control their preferred folding direction. A shim embedded across a hinge produces an asymmetric torque–angle response that favors either mountain or valley folding. Switching the shim between its two stable states reverses this response, allowing the folding direction of each hinge to be reprogrammed after fabrication. By independently controlling the states and geometries of the shims, we enable a single origami sheet to access multiple folding branches and transform into prescribed three-dimensional shapes. We further introduce self-switching hinges in which folding causes the shims to snap between their stable states. These elements allow the sheet to reprogram its folding pathway  under    global mechanical inputs applied to the boundaries. Our approach embeds both shape selection and transition rules directly within the mechanics of the hinges, providing a versatile framework for creating multifunctional, deployable, and reconfigurable structures.}


\maketitle
\clearpage

\section{Introduction}\label{Intro}

Origami, the traditional Sino-Japanese art of folding flat sheets into complex three-dimensional shapes, has evolved into a versatile engineering platform for reconfigurable and deployable structures, with applications in aerospace, robotics, medicine, and architecture \cite{ misseroni2024origami,meloni2021engineering,reis_transforming_2015,johnson2017fabricating}. Origami structures consist of panels connected by flexible hinges. Starting from a flat configuration, they can fold along multiple branches, each leading to a distinct three-dimensional shape \cite{ginepro2014counting,dieleman_jigsaw_2020,chen_branches_2018}. Accessing a desired branch requires biasing each hinge toward either a mountain or a valley fold. From a purely geometric perspective, a basic question is that of reachability, i.e., can one fold an arbitrary shape from a flat sheet of paper, a question that has been addressed using a range of different approaches \cite{lang2011origami}, and most recently via both global optimization and local marching methods with constructive proofs \cite{dudte2016programming, dudte2021additive}. From a mechanical perspective, this has been traditionally implemented by creasing and plastically deforming the hinges in the prescribed directions \cite{rao_fold_2013}, but this process permanently and irreversibly alters the sheet, making it difficult to access alternative folding branches after the initial fold. Reversible control over the preferred folding direction of individual hinges could greatly expand the range of shapes and folding pathways accessible to a single origami structure, enabling multiple functionalities to be encoded within the same sheet. This potential has been demonstrated by embedding actuators at individual hinges, allowing an origami sheet to be programmed to fold into multiple target shapes \cite{hawkes_programmable_2010}.

Beyond active control, multistability offers a passive strategy for switching an origami sheet among multiple stable mountain–valley configurations. By exploiting geometric incompatibilities between folding branches, multistable origami structures can stably access distinct configurations without the need for electronics. This approach has enabled deployable inflatable shelters \cite{melancon2021multistable}, Miura--ori sheets with reprogrammable deformation modes \cite{silverberg_using_2014}, inflatable actuators with multiple deformation pathways \cite{melancon2022inflatable}, and mechanical logic elements \cite{treml_origami_2018}. However, multistability arises naturally in only a limited class of origami patterns.
 
To extend multistability to a broader range of origami patterns, recent studies have introduced multistability through hinge mechanics rather than relying solely on the geometry of the fold pattern. For example, multistability has been demonstrated in four-vertex origami unit cells by incorporating torsional springs with nonzero rest angles at the hinges \cite{waitukaitis_origami_2015} and in non-Euclidean origami by allowing the creases to stretch \cite{addis_connecting_2023}. Bistable elements embedded within the hinges have also been used to enhance the stiffness of origami structures \cite{michalaros_stiffening_2025}, while prestretched springs across hinges have enabled double-symmetric four-vertex origami to exhibit stable configurations along multiple folding branches \cite{iniguez-rabago_rigid_2022}. Although these approaches expand the number of accessible stable configurations, they offer limited control over switching among them. 

Beyond origami, embedding switchable bistable elements within elastic structures has emerged as an effective strategy for controlled switching between shapes and reprogramming global mechanical behavior through local inputs. Snapping shells have been used to tune structural stiffness \cite{chen_reprogrammable_2021}, while buckling domes integrated into elastic sheets have enabled shape-morphing systems that transition among multiple stable configurations, with applications in robotics \cite{faber2020dome,liu_snap-induced_2023,osorio2023manta}. Similarly, bistable buckled strips have been exploited to realize reconfigurable metastructures for space applications \cite{lalisani_snap-through_nodate} and morphing drones \cite{risso_highly_2022,girardi_multistable_2025}. 

Motivated by these advances, we integrate bistable buckled beams into origami hinges to create a versatile platform for programmable shape morphing. We first show that a buckled shim embedded across a hinge generates an asymmetric torque–angle response, biasing the hinge toward either a mountain or a valley fold. Switching the shim between its two stable states reverses this bias, allowing the preferred folding direction to be reprogrammed after fabrication. By independently controlling the state of each shim, a single origami sheet can access multiple folding branches, while tuning the shim geometry enables these branches to produce prescribed target shapes. We further design the shims to snap passively as the hinges rotate, allowing the structure to reprogram itself sequentially in response to a global mechanical input. By incorporating these switching mechanisms into shape-changing sheets, we encode prescribed transitions between distinct configurations within a single structure, providing a foundation for new classes of deployable and reconfigurable systems.

\section{Results}\label{Results}
\subsection{Characterization of a bistable buckled hinge} 
\begin{figure}[hb]
\centering
\includegraphics[width=\linewidth]{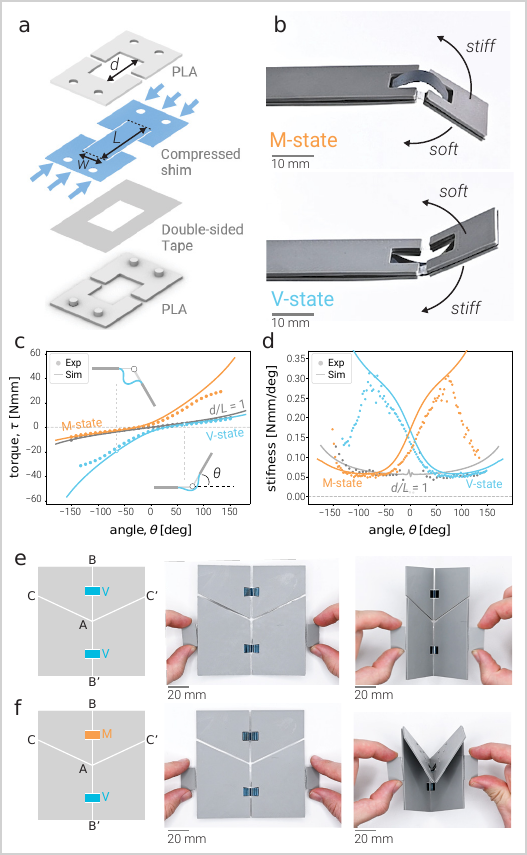}
\caption{(a) Exploded schematic of an origami hinge incorporating a bistable buckled shim. (b) Photographs of a fabricated hinge in its two stable states. (c) Torque--angle and (d) stiffness--angle responses of the hinge in each stable state. (e,f) Folding of a four-vertex origami structure with two shims buckled in the same direction (e) or in opposite directions (f).}\label{fig1}
\end{figure}

The origami sheets in this work are assembled by sandwiching double-sided adhesive tape between two 0.8 mm-thick 3D-printed PLA panels (Fig.~\ref{fig1}a). The PLA panels are spaced 1.7 mm apart and serve as the rigid faces of the origami, while the tape forms a compliant hinge between the panels. At each hinge we create a rectangular gap with length $d = 15$ mm and insert a spring-steel shim of length $L$, width $w$, thickness $t =0.05$ mm and Young's modulus $E = 169.6$ GPa.
For $L>d$, the shim must be compressed to fit within the gap and consequently buckles into an arched shape. The buckled shim exhibits bistability \cite{cazzolli_snapping_2019,gomez_critical_2017,plaut_vibration_2009} and can be pressed to switch between two stable configurations, curved upward or downward. We refer to these configurations as the mountain (M) and valley (V) states, respectively, as they induce a mountain and valley fold at the hinge (Fig.~\ref{fig1}b). 

We characterize the mechanical response of the bistable buckled hinges by measuring their torque--rotation profiles using a rheometer (see SI S1.2). 
We consider hinges with $d/L = 0.9$ and $w = 8$~mm in both their mountain~(M) and valley~(V) states and  compared them to those with a flat, unbuckled shim (i.e. $d = L$ and $w = 8$~mm). 
Fig.~\ref{fig1}c displays the measured torque, $\tau$, as a function of the applied angle, $\theta$, between the two panels, where $\theta = 0$ corresponds to a flat fold, $\theta > 0$ to a valley fold, and $\theta < 0$ to a mountain fold (see inset in Fig.~\ref{fig1}c).
Furthermore, in Fig.~\ref{fig1}d we show the evolution of the corresponding rotational stiffness, $k_\theta$, calculated as $k_\theta=d\tau/d\theta$.  For the hinges with the unbuckled shim, the torque $\tau$ increases approximately linearly with the rotation angle $\theta$ over the investigated range, resulting in a nearly constant rotational stiffness of $k_\theta \approx \text{55 N}\mu\text{m/deg}$. In contrast, the hinges with the buckled shim exhibit a markedly asymmetric response to positive and negative rotations. The torque follows a trend similar to that of the unbuckled shim when the hinge is bent along the curvature of the shim to increase $\theta$ for a shim in the M-state and to decrease $\theta$ for a shim in the V-state. This corresponds to the soft direction indicated in Fig.~\ref{fig1}b. However, the torque rises more rapidly in the opposite direction against the curvature of the shim, resulting in a substantially higher rotational stiffness. This direction is denoted as the stiff direction in Fig.~\ref{fig1}b.  In particular, for the shim in the M-state, the rotational stiffness at $\theta=65^\circ$ is 4.3 times higher than that at $\theta=-65^\circ$. Conversely, for the shim in the V-state, this asymmetry is reversed, with $k_\theta$ at $\theta=65^\circ$ being 4.4 times lower than at $\theta=-65^\circ$. This is attributed to the higher curvature in the shim for the stiff bending direction than for the soft bending direction (see insets Fig.~\ref{fig1}c).

The behavior of the shim can be captured using a simple one-dimensional elastic beam model that considers only the metallic shim and neglects the effects of the panels and adhesive tape (see SI S1.3). As shown in Fig.~\ref{fig1}c,d, the model accurately reproduces the experimentally measured torque--rotation and stiffness responses. The agreement between model and experiments indicates that the mechanical behavior of the hinge is dominated by the deformation of the metallic shim, while the contributions of the panels and adhesive tape are comparatively small. For large rotations in the stiff bending direction we see deviation from the simulated curves, which are attributed to plasticity. In this work we avoid these large rotations so that the response of the origami sheets can be assumed to be fully elastic.

The stiffness asymmetry introduced by the buckled metallic shims can be exploited to bias origami structures into distinct deformation branches upon folding from the flat state. As an example, we consider the fundamental four-vertex building block of the Miura--ori pattern and incorporate two metallic shims with $d/L=0.9$ along two of its hinges (Fig.~\ref{fig1}e,f). When the two panels connected by hinge AB$'$ are pushed together, the resulting folding behavior depends on the states of the buckled shims. If both shims are in the same state, the structure folds symmetrically about the BB$'$ hinge, producing the configuration shown in Fig.~\ref{fig1}e. In contrast, if the shims are in different states, hinges AB and AB$'$ are biased to fold in opposite directions. To accommodate this, the sheet folds along the diagonal hinge AC, resulting in the configuration shown in Fig.~\ref{fig1}f.

\subsection{Combining multiple shims on a single hinge} 

\begin{figure*}[t]
\centering
\includegraphics[width=\textwidth]{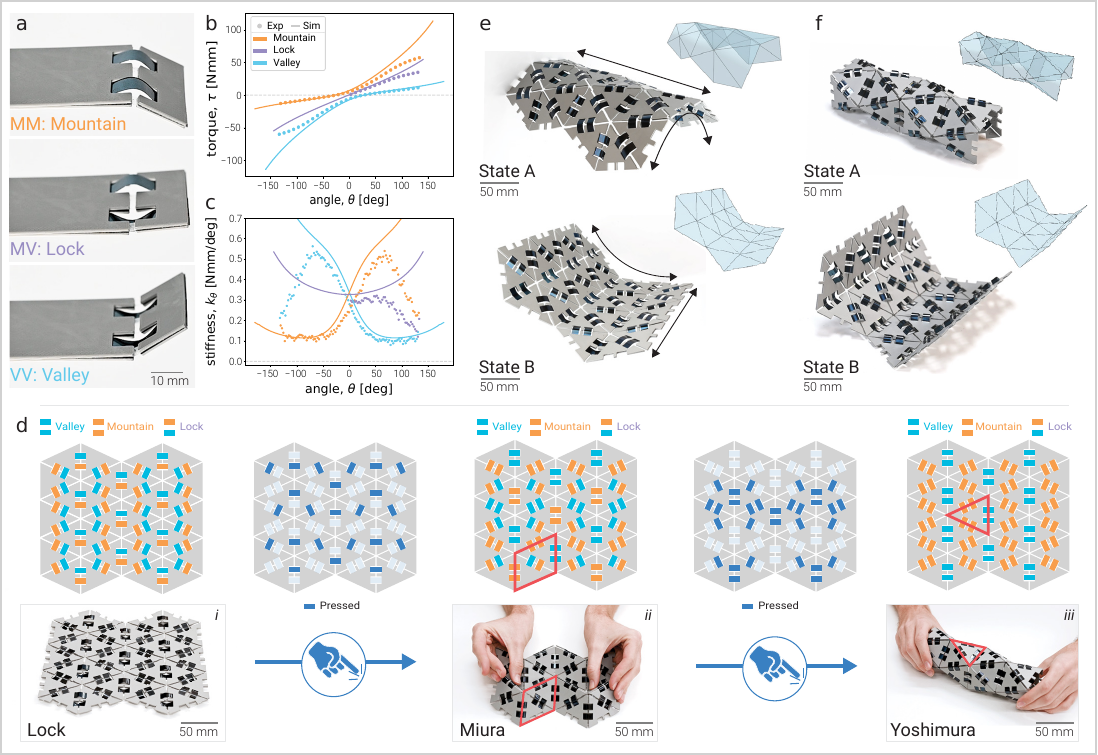}
\caption{(a) Photographs of a fabricated hinge containing two parallel shims. (b) Torque--angle and (c) stiffness--angle responses of the hinge in the three supported  stable states of the buckled shims. (d) Reconfiguration of a triangulated sheet with bistable hinges among the locked-flat, Miura, and Yoshimura configurations, together with diagrams of the corresponding encoded shim states. (e,f) Stable Yoshimura configurations of triangulated sheets with bistable hinges for shim geometries of $d/L=0.9$ (e) and $d/L=0.75$ (f), shown through experimental images and numerical predictions obtained using the spring-based model.}\label{fig2}
\end{figure*}

The range of accessible hinge states can be further expanded by placing a second buckled shim across a hinge (Figs.~\ref{fig2}a--c).  In addition to the mountain~(MM) and valley~(VV) configurations, the hinge can be programmed into a locked~(MV) configuration.  In this MV state, the rotational stiffness is high for both positive and negative rotations, effectively locking the hinge in its flat configuration.  Consequently, degrees of freedom can be selectively removed and restored in the origami sheet.  This allows fold patterns with completely different topologies to be reversibly programmed by simply pressing specific shims.

To demonstrate this concept, we consider a tessellation of triangular panels (Fig.~\ref{fig2}d). 
Without the metal shims, this structure is highly floppy and difficult to fold into prescribed shapes due to its large number of degrees of freedom. 
However, by incorporating two metallic shims at each hinge and setting all of them to their MV state, the sheet becomes rigid and is constrained from folding (Fig.~\ref{fig2}d(i)). 
A sequence of programmatic presses can then transform the sheet from this flat state into two distinct origami fold patterns. 
Specifically, changing the state of 23 shims programs the Miura--Ori pattern, which folds into a planar corrugated configuration (Fig.~\ref{fig2}d(ii)). 
Switching an additional 30 shims subsequently reprograms the structure into the Yoshimura pattern, causing it to fold into a cylindrical shape (Fig.~\ref{fig2}d(iii)). 
Notably, these two configurations require fundamentally different unit-cell topologies. The Miura--Ori pattern is a quad based topology with four hinges folding at each vertex, while the Yoshimura is a triangular fold pattern with six hinges folding at each vertex, as highlighted by the red outlines in Figs.~\ref{fig2}d(ii) and \ref{fig2}d(iii). 
An additional demonstration showcasing reprogramming between the Miura--Ori and Kresling patterns is provided in the Supplementary Information (Section S2.3).

Interestingly, for a given configuration of the buckled shims, the folded origami sheets can exhibit multistability. For example, when the buckled shims are programmed to realize the Yoshimura pattern, the sheet possesses two distinct stable equilibrium configurations upon release, denoted as states A and B in Fig.~\ref{fig2}e. The sheet can be switched reversibly between these two stable states without pressing any of the buckled shims (Video S2). As shown in Fig.~\ref{fig1}b, each hinge has a non-zero rest angle (e.g. $\theta \neq 0$ when $\tau=0$). However, the geometric compatibility constraints imposed by the origami sheet prevent all hinges from simultaneously reaching their preferred rest angles. This frustration generates an elastic energy landscape with multiple local minima, giving rise to the observed multistable behavior.  To find the stable states of the sheets we developed a spring based model, inspired by previous efforts \cite{liu2017nonlinear}, where gradient information is used to find vertex positions minimizing a sheet energy. Simulation details are available in the work of Liu et al. \cite{liu2026basin}. Through this we predict the shapes shown in Fig.~\ref{fig2}e and f as stable energy minima, which are shown as insets. In this way stable configurations can be found without the need to fabricate and manually manipulate the sheet. As an example, the model predicted an additional undulating stable state for the sheet in the Yoshimura configuration which was later experimentally verified (See SI S2.1).

\begin{figure}[H]
\centering
\includegraphics[width=\linewidth]{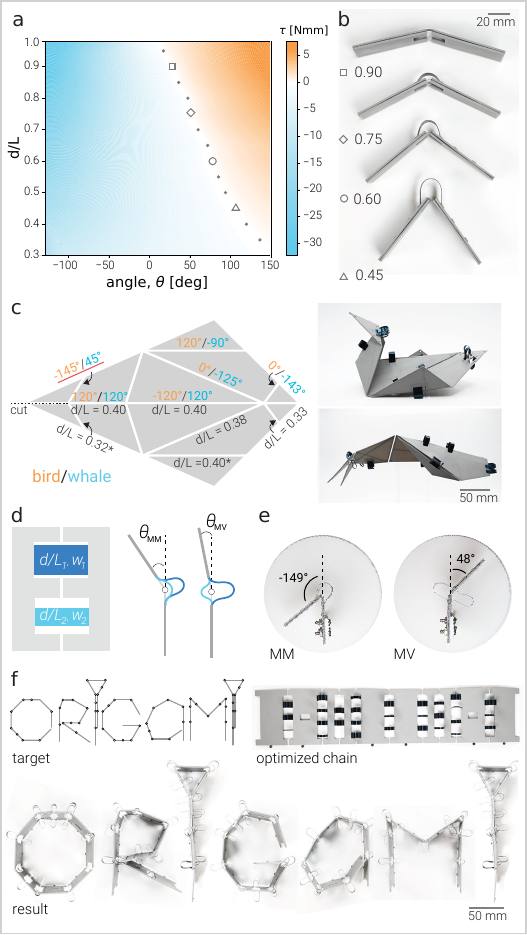}
\caption{(a) Experimentally measured torque--angle responses linearly interpolated to express the torque as a continuous function of $\theta$ and $d/L$. (b) Photographs of hinges containing shims of different lengths int heir rest configuration. (c) Hinge pattern designed to fold into bird and whale shapes, together with photographs of the fabricated origami in the two shapes. The upper half of the schematic indicates the target hinge angles, while the lower half shows the corresponding shim geometries. (d) Schematic of a hinge containing two parallel shims with different geometries. (e) Photographs of a hinge designed with rest angles $\theta^{MV}_0=45^\circ$ and $\theta^{MM}_0=-145^\circ$. (f) Chain of hinged panels programmed to form the letters in ``ORIGAMI''.}\label{fig3}
\end{figure}

The shape of these stable configurations can be tuned by varying  $d/L$. 
Fig.~\ref{fig2}f shows the two stable configurations of a sheet with metallic shims with $d/L = 0.75$ programmed into the Yoshimura pattern. Both stable states exhibit a higher curvature than their counterparts obtained with $d/L = 0.9$, indicating  that the ratio $d/L$  controls the rest angle of the hinges.
 To quantify this effect, we measured the torque--rotation response of hinges programmed in the M state for $d/L \in [0.30,1.00]$ in increments of 0.05 using a rheometer. The measurements were linearly interpolated to obtain the torque as a continuous function of both $\theta$ and $d/L$ (Fig.~\ref{fig3}a). We find that the hinge rest angle $\theta_0$ increases with decreasing $d/L$ (Fig.~\ref{fig3}b). Collectively, these measurements demonstrate that selecting an appropriate $d/L$ enables programming of hinge rest angles spanning the full range from $0^\circ$ to $180^\circ$.

 This ability to program the rest angle $\theta_0$ can be exploited to design origami sheets that support two distinct target shapes in their stress-free configurations. 
Fig.~\ref{fig3}c shows the target hinge angles required to fold a kite-shaped sheet into either a bird (orange) or a whale (blue).  These target angles are subsequently mapped to specific shim lengths using the torque--angle profiles shown in Fig.~\ref{fig3}a. 
However, certain hinges must accommodate two different rest angles to fold the two shapes. 
To achieve this, we adopt the dual-shim strategy introduced in Fig.~\ref{fig2}, while allowing the geometric ratio $d/L$ and the bending stiffness $B = Ewt^3/12$ of each shim to vary independently. 
The geometric ratio $d/L$ dictates the individual rest angle of a shim and the bending stiffness $B$---tuned by adjusting the shim width $w$---linearly scales the corresponding torque--angle response. To design a hinge with two prescribed rest angles, $\theta_{0}^{MM}=-\theta_{0}^{VV}$ for the MM and VV states and $\theta_{0}^{MV}$ for the MV state, we solve for the lengths $L_i$ and bending stiffnesses $B_i$ ($i=1,2$) of the two shims that make the total torque vanish at the desired equilibrium configurations:

\begin{equation}
\label{EqB}
\begin{aligned}
\frac{B_1}{\tilde B}\tau\left(\frac{d}{L_1},\theta_0^{MM}\right)+
\frac{B_2}{\tilde B}\tau\left(\frac{d}{L_2},\theta_0^{MM}\right) &= 0, \\
\frac{B_1}{\tilde B}\tau\left(\frac{d}{L_1},\theta_0^{MV}\right)-
\frac{B_2}{\tilde B}\tau\left(\frac{d}{L_2},-\theta_0^{MV}\right) &= 0.
\end{aligned}
\end{equation}
Here, $\tilde B$ is the bending stiffness of the reference shim used to generate the torque--angle curves in Fig.~\ref{fig3}a.
Since the torque scales linearly with bending stiffness, the equilibrium condition depends only on the stiffness ratio $B_2/B_1$. We therefore solve the governing equations for $B_2/B_1$ rather than for $B_1$ and $B_2$ separately. As an example case, we consider the leftmost hinge in Fig.~\ref{fig3}c, which requires a rest angle of $\theta_0^{MM}=-145^\circ$ for the bird  and $\theta_0^{MV}=45^\circ$ for the whale. When we impose the simplifying constraint $L_1=L_2$,  solving Eq.~(\ref{EqB}) yields $d/L_1=d/L_2=0.32$ and $B_1/B_2=0.385$. 
To achieve this required bending stiffness ratio, we fabricate a hinge featuring two metallic shims with widths $w_1 = 8$~mm and $w_2 = 3.08$~mm. 
For this hinge, we measure rest angles of $\theta_{0}^{\text{MM}} = -149^\circ$ and $\theta_{0}^{\text{MV}} = 48^\circ$. 
These experimental values deviate from the target angles by only $4^\circ$ and $3^\circ$, respectively (Fig.~\ref{fig3}e), demonstrating the accuracy of the proposed design framework despite minor discrepancies likely caused by fabrication imperfections. 
By applying this identical design procedure to all hinges, we successfully realize an origami sheet capable of folding into both a bird and a whale (Fig.~\ref{fig3}c), as well as a second sheet that folds into a box and a boat (see SI S3.3).

By incorporating $N$ distinct buckled shims into a hinge, the number of accessible rest angles can be increased to $2^{N-1}$. Fig.~\ref{fig3}f shows a one-dimensional array of 10 panels connected by hinges containing up to four buckled shims. The hinges are programmed to support up to five distinct target rest angles, enabling the array to fold into configurations representing all six distinct letters of the word ORIGAMI. Experiments confirm that each letter is clearly recognizable with a deviation from the targeted angles of $4\pm4^\circ$. Additional design details and characterization are provided in the SI S3.4.

\begin{figure*}[hb!]
\centering
\includegraphics[width=\textwidth]{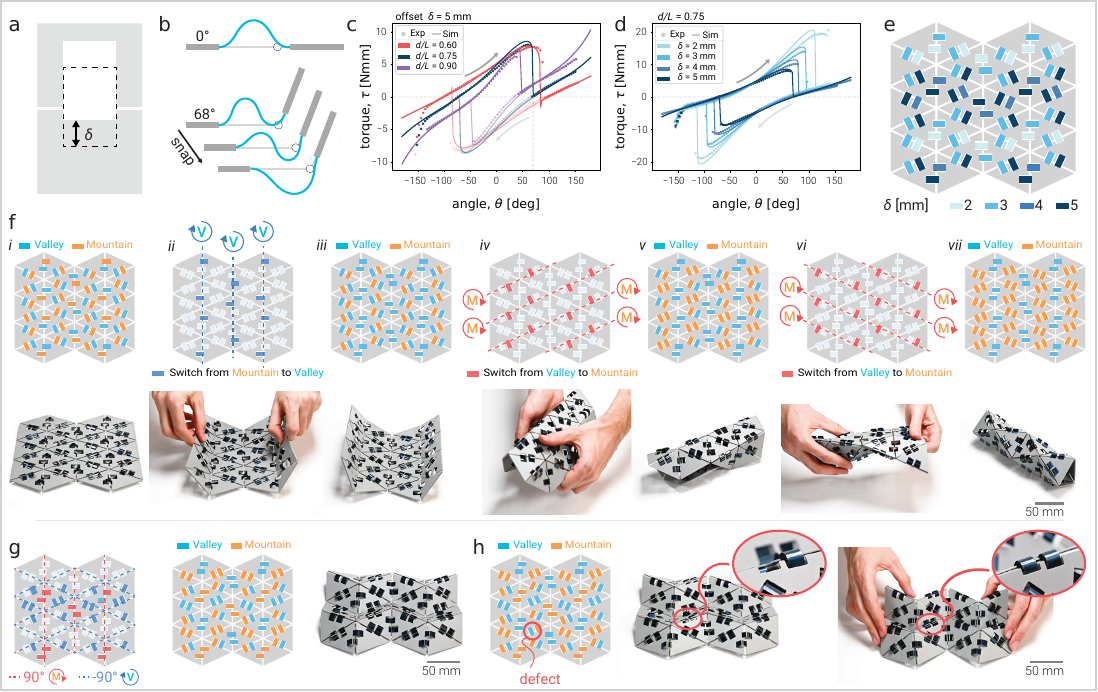}
\caption{(a) Schematic of a hinge with the shim offset from the hinge axis by a distance $\delta$. (b) Simulated snap-through of a shim with $d/L=0.75$ and $\delta=5$ mm during hinge rotation. (c) Torque--angle responses for shims with $\delta=5$ mm and varying $d/L$. (d) Torque--angle responses for shims with $d/L=0.75$ and varying $\delta$. (e) Schematic of a triangulated sheet containing offset shims. (f) Reconfiguration of the sheet in (e) from a locked-flat state to the Yoshimura configuration. Dark blue and red dashed lines indicate folds at which the buckled shims switch states in response to an input applied at the boundaries. (g) Folding sequence and final shim states associated with reconfiguring the sheet from the Yoshimura to the Miura configuration. (h) Passive accommodation of a defect in the Miura configuration.}\label{fig4}
\end{figure*}

\subsection{Reprogramming   patterns with global inputs}

The examples in Figs.~\ref{fig2} and \ref{fig3} demonstrate that locally pressing the buckled shims can reprogram the shape of an origami sheet. However, individually actuating each shim results in prohibitively long reprogramming times for larger origami structures. We therefore explore passive switching strategies that cause the shims to transition between their stable states in response to global mechanical inputs, eliminating the need for localized actuation of individual shims.

We start by harnessing a well known snap-through instability that is triggered when the ends of a buckled shim are rotated in the shim's stiff direction, causing the shim to transition to its second stable state \cite{cazzolli_snapping_2019}. When embedded within an origami hinge, the ends of the shim undergo a similar rotation during folding. However, for the shim placement considered thus far, snap-through is suppressed upon rotation. As the hinge folds, the distance between the two ends of the shim decreases, increasing the energy barrier associated with snap-through.
 This can be overcome by offsetting the shim from the hinge axis by a distance $\delta$, as shown in Fig.~\ref{fig4}a. Fig.~\ref{fig4}b shows numerical snapshots of a shim initially in the M-state with $d/L=0.75$ and $\delta=5$ mm. The offset of the shim reduces the end-to-end distance of the shim upon rotation in the stiff direction, which allows a snap-through at a critical angle of $\theta_s=68^\circ$. This instability is accompanied by a sharp drop in the torque--angle response, as shown in Figs.~\ref{fig4}c,d for shims with different combinations of $d/L$ and $\delta$. These results show that the critical snapping angle $\theta_s$ decreases with decreasing $d/L$ and increasing  $\delta$, providing a means to tune the switching threshold. Furthermore, they indicate that the rest angle $\theta_0$ is only marginally affected by $\delta$.

By incorporating shims with $\delta\neq0$ into an origami sheet, we can realize origami structures that can be reprogrammed between different folding patterns through mechanical manipulation of their boundaries, eliminating the need to directly actuate individual shims. As an example, Fig.~\ref{fig4}e shows the triangular tessellation introduced in Fig.~\ref{fig2}, where all hinges contain shims with $d/L=0.75$ and $w=8$ mm that are offset by $\delta\in[2,5]$ mm.
We begin with all hinges in the MV state so that the sheet is locked in the flat state (Fig.~\ref{fig4}f(i)). A mechanical input is then applied to rotate the three vertical fold lines highlighted by the blue dashed lines in Fig.~\ref{fig4}f(ii) in the valley direction. This rotation causes all shims in the mountain state to snap to their valley state, transforming the initially flat sheet into a curved configuration (Fig.~\ref{fig4}f(iii)). Next, a second mechanical input is applied to rotate the diagonal fold lines highlighted by the red dashed lines in Fig.~\ref{fig4}f(iv)-(vi) in the mountain direction. This causes the shims that were initially programmed in the valley state along these folds to snap into the mountain state, reprogramming the fold pattern into the Yoshimura configuration with its characteristic cylindrical shape Fig.~\ref{fig4}f(vii).

Reprogramming the sheet from the Yoshimura to the Miura--Ori pattern is more challenging because only some hinges along a fold line must switch between mountain and valley states, while others must remain unchanged to produce, for example, the locked MV hinges. To achieve this selective switching, we exploit the dependence of  $\theta_s$ on $\delta$. Rotating a fold line by $90^\circ$ in the stiff direction causes only the shims with $\delta=4$ and $5$ mm to snap, while the shims with $\delta=2$ and $3$ mm remain in their original state, since $\theta_s^{\delta=3 mm}>90^\circ>\theta_s^{\delta=4 mm}$. Fig.~\ref{fig4}g illustrates the fold operations used to reprogram the sheet from the Yoshimura configuration to the Miura--Ori pattern.
Note that the sheet shown in Fig.~\ref{fig4}e was specifically designed to switch between the locked, Miura--Ori, and Yoshimura configurations with a small number of folding operations. 
Therefore, the sheet does not support every possible fold pattern. Enabling arbitrary reprogramming would require a greater diversity of shim offsets along each continuous fold line, thereby providing additional independently addressable snapping thresholds. Finally, we note that, similar to conventional paper origami, the programmed origami retains a memory of the imposed folding direction. When the hinges are rotated beyond $\theta_s$, the sheet relaxes to an equilibrium that approximates the imposed configuration, as is shown in Fig.~\ref{fig4}f. Furthermore, the programmed fold pattern exhibits a degree of self-correction. As an example, we introduce a defect into a Miura--Ori sheet by switching a single shim from the mountain to the valley state. Upon compression, the surrounding fold pattern constrains the defective hinge to follow the prescribed Miura--Ori kinematics. As a result, the hinge snaps back to the M state, restoring the correct buckled-shim configuration underlying the Miura--Ori fold pattern (Fig.~\ref{fig4}h).

\begin{figure}[t]
\centering
\includegraphics[width=\linewidth]{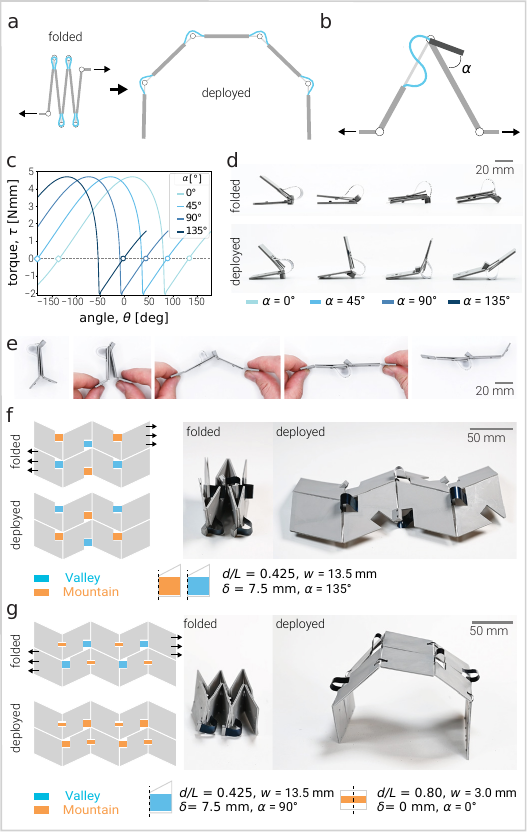}
\caption{(a) Concept of a deployable structure. (b) Schematic of a hinge with  $\delta=7.5$ mm and  the end
located near the hinge connected to a rigid clamp oriented
at an angle $\alpha$ relative to the panels. (c) Torque--angle responses and (d) photographs of deployable hinges with $d/L=0.425$ and $\delta=7.5$ mm for different values of $\alpha$. (e) Snapshots of a hinge with $d/L=0.425$, $\delta=7.5$ mm and $\alpha=135^\circ$ transitioning from a compact to a deployed state. (f) Deployment of 4$\times$2 Miura–Ori sheet from its compact configuration. (g) Deployment of a  5$\times$2 Miura–Ori sheet from its compact   configuration to an arched configuration.}\label{fig5}
\end{figure}

Shims with $\delta \neq 0$ can also be used to create deployable origami structures that transform from a compact folded state into an expanded configuration when subjected to a tensile load at their boundaries (Fig.~\ref{fig5}a). This functionality requires hinges that snap between two distinct rest angles: one near $\pm180^\circ$, corresponding to the compact state, and another of smaller magnitude, corresponding to the deployed state.
To achieve this response, we set the shim offset to a value of $\delta=d/2=7.5$ mm. This configuration places the buckled shim entirely on one side of the hinge and minimizes its critical snapping angle, $\theta_s$. Moreover, rather than attaching both ends of the shim directly to the adjacent panels, we connect the end located near the hinge to a rigid clamp oriented at an angle $\alpha$ relative to the panels (highlighted in dark gray in Fig.~\ref{fig5}b). As a result, the hinge exhibits two rest angles, $-\theta_0+\alpha$ and $\theta_0+\alpha$, where $\theta_0$ is the rest angle of the shim when $\alpha=0$.
Figs.~\ref{fig5}c,d show   the evolution of the torque--angle response and the rest angles as a function on $\alpha$ for shims with $d/L=0.425$ and $\theta_0=133^\circ$, for which $\theta_s=83^\circ$.  The results indicate that by varying the clamp angle $\alpha$, a wide range of hinge rest-angle pairs can  be realized with contact between the panels preventing rest angles exceeding $\pm180^\circ$. In particular, when $\alpha>47^\circ$, one of the rest angles saturates at $-180^\circ$, producing a fully folded configuration, while the second rest angle can be continuously tuned to any value below $90^\circ$.

Hinges with $\alpha>\theta_s$ are particularly attractive for deployable structures because they snap before reaching a flat configuration at $\theta = 0$. Fig.~\ref{fig5}e shows snapshots of an hinge with $\theta_0=-55^\circ$, $\alpha=135^\circ$. In its compact configuration, the hinge is fully folded at $-180^\circ$. As the structure is pulled open, the hinge rotates until it reaches the snapping angle, where it spontaneously transitions to its second stable state with a flat rest angle of approximately $-2^\circ$. Consequently, origami structures incorporating such hinges can be deployed simply by applying a tensile load at their boundaries.
As a demonstration, we integrate these hinges into a $4\times2$ Miura--Ori sheet (Fig.~\ref{fig5}f). In its compact configuration, the panels stack on top of one another. Applying a tensile load at the boundaries causes the hinges to rotate beyond their critical snapping angle, triggering their transition to the deployed state. This expands the compact stack into an expanded almost flat Miura--Ori pattern. During this transformation, all shims snap, reversing the assignment of every crease from mountain to valley and vice versa.

More complex shape transformations can be achieved by selectively combining snapping and non-snapping buckled shims within a single origami sheet. As a demonstration, we extend the $4\times2$ Miura--Ori design to a $5\times2$ sheet that deploys into an arch (Fig.~\ref{fig5}g). To achieve this, along one zigzag path, we replace the snapping shims with centered, non-snapping shims characterized by $\delta=0$, $w=3.0$~mm, $d/L=0.80$, and $t=0.05$~mm, while the clamp angle of the remaining snapping shims is changed to $\alpha = 90$. When a tensile load is applied at the boundaries of the sheet, the snapping hinges transition to their second stable state, while the non-snapping shims remain in their initial state. Consequently, all vertical hinges adopt a mountain-fold configuration with a rest angle of approximately $-45^\circ$, producing the arched structure shown in Fig.~\ref{fig5}g. A more detailed description of these deployable structures in given in SI S4 and full deployment is shown in video S5.


\begin{figure*}[b]
\centering
\includegraphics[width=\textwidth]{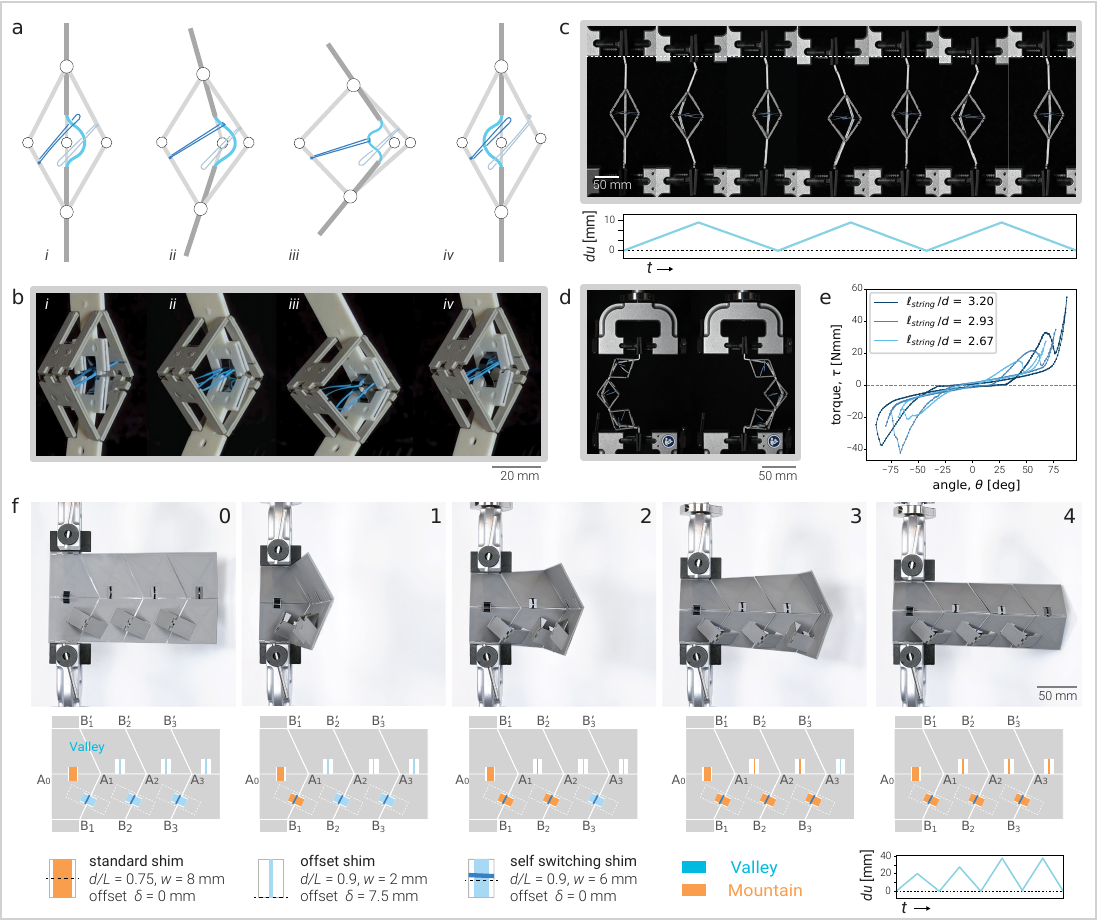}
\caption{(a) Schematic and (b) photographs of a self-switching hinge.(c) Snapshots of a self-switching hinge subjected to three cycles of compression. (d) Snapshots of a chain with three self-switching hinges subjected to two cycles of compression. (e) Torque–angle profiles of self switching hinges with different string lengths. (f) Snapshots and diagrams of a $4 \times 2$ array of Miura--ori panels  designed to shift its folding location progressively with each compression cycle, thereby mechanically counting the number of loading cycles.}\label{fig6}
\end{figure*}

The deployable structures shown in Fig.~\ref{fig5} are deployed by applying a linear tensile load at their boundaries, significantly reducing the actuation complexity compared to the structures shown in Figs.~\ref{fig2},~\ref{fig3}  and~\ref{fig4}. However, resetting them to their compact folded configuration is more challenging, as it requires individually addressing each hinge to reprogram it to its initial folded state. While this behavior is desirable for applications in which the deployed configuration should be retained robustly, it is less suitable for systems that must repeatedly switch between multiple shapes.
We therefore seek to engineer origami sheets that undergo sequential reconfiguration between multiple shapes under a single repetitive mechanical input. To achieve this, rather than using shims that snap when rotated in their stiff direction, we engineer shims that snap when rotated in their soft direction. These hinges will have a passive bias to rotate in the direction that will change the state of the shim. Applying compression at the boundaries of sheets with these self-switching hinges is therefore likely to reconfigure their state and folding behavior.

To enable this behavior, we introduce two opposing strings that drive the buckled shim between its two stable states as the hinge rotates (Figs.~\ref{fig6}a,b). The string is looped around a buckled shim with $d/L = 0.9$, $w = 6$ mm and $\delta = 0$ mm and has its ends connected to a pair of auxiliary hinged panels (highlighted in light gray in Fig.~\ref{fig6}a), which are attached to the main panels on either side of the hinge. Initially, the hinge rotates in the soft direction of  the buckled shim. As it rotates, the attachment points of the string on the side opposite the direction in which the shim bows move away from the shim, while those on the side toward which the shim bows move closer to it. Consequently, the string on the side opposite the   buckling direction of the shim becomes taut, whereas the other string remains slack. Once the hinge reaches a critical rotation, the tension in the taut string causes the shim to snap into its other stable state (Figs.~\ref{fig6}a,b(iii)). The shim remains in this state when the hinge returns to $\theta=0^\circ$ (Figs.~\ref{fig6}a,b(iv)), thereby reversing  preferred folding direction of the hinge. When the hinge is rotated again, it therefore folds in the opposite direction, causing the other string to become taut and eventually snap the shim back to its original state. In this way, the hinge passively reverses its preferred folding direction after each applied rotation (Fig.~\ref{fig6}c). The change in folding direction induced by the strings becomes more apparent when multiple hinges are connected in series. Fig.~\ref{fig6}d shows a chain of three self-switching hinges subjected to two consecutive compression cycles. During the first cycle, the chain forms a half-sinusoidal profile that bends to the left; during the second, it bends to the right. Finally, the angle at which the string causes the buckled shim to snap can be tuned by varying the string length, $\ell_{\text{string}}$. Fig.~\ref{fig6}e shows the torque--angle responses of self-switching hinges for different values of $\ell_{\text{string}}$. Increasing the string length delays snapping because the hinge must rotate through a larger angle before a string becomes taut.

Next, we combine the self-switching hinges of Fig.~\ref{fig6}a-e with the offset snapping hinges introduced in Fig.~\ref{fig4} to create origami sheets that undergo complex sequential deformations under repeated compression. Fig.~\ref{fig6}f shows a $4\times2$ array of Miura--ori panels designed to fold progressively at different locations with each compression cycle, effectively ``counting'' the number of times the sheet has been loaded and changing its shape accordingly. 
Upon loading, the sheet follows the folding pathway that minimizes the elastic energy stored in the hinges and, consequently, minimizes the number of hinges that bend in the stiff direction. Guided by this principle we realize the counter by placing three different buckled shims across the hinges (Fig.~\ref{fig6}f(0)): ($i$)  three self-switching shims with $d/L=0.9$, $w=6$~mm, $\delta=0$~mm, and $\ell_{\text{string}}/d=3.2$ along the lower diagonal hinges $\mathrm{A}_i\mathrm{B}_i$; ($ii$) a non-snapping shim with $d/L=0.75$, $w=8$~mm, and $\delta=0$~mm in the leftmost horizontal hinge, $\mathrm{A}_0\mathrm{A}_1$; and ($iii$) three  offset snapping shims with $d/L=0.9$, $w=2$~mm, and $\delta=7.5$~mm in the remaining horizontal hinges $\mathrm{A}_i\mathrm{A}_{i+1}$ (with $i$=1, 2 and 3).  Initially, the non-snapping shim at the hinge $\mathrm{A}_0\mathrm{A}_1$ is set to the M-state, while all other shims are initialized in the V-state.

When a compressive input is applied to the two panels connected by hinge $\mathrm{A}_0\mathrm{A}_1$, the hinges $\mathrm{A}_0\mathrm{A}_1$, $\mathrm{A}_1\mathrm{B}_1$, and $\mathrm{A}_1\mathrm{B}'_1$ fold in their preferred directions, while all other hinges remain flat. As the structure folds, the string causes the self-switching shim at $\mathrm{A}_1\mathrm{B}_1$ to snap to the M-state (Fig.~\ref{fig6}f(1)). Upon removal of the load, the sheet returns to its flat configuration while retaining the switched state of the shim at $\mathrm{A}_1\mathrm{B}_1$.

During the next compression cycle, the switched state of the buckled shim at $\mathrm{A}_1\mathrm{B}_1$ makes rotation of this hinge energetically unfavorable because it would require bending the shim in its stiff direction. Folding therefore advances to the next vertex, where hinge $\mathrm{A}_2\mathrm{B}_2$ rotates in its soft direction (Fig.~\ref{fig6}f(2)). Note that this deformation requires the offset shim at the horizontal hinge $\mathrm{A}_1\mathrm{A}_2$ to bend in its stiff direction. To favor folding of the second vertex over refolding of the first, we design the offset shim to be narrow in width so that its bending in the stiff direction requires less energy than the previously switched self-switching shim at $\mathrm{A}_1\mathrm{B}_1$. As hinge $\mathrm{A}_1\mathrm{A}_2$ rotates, the offset shim snaps and reverses its preferred folding direction, thereby eliminating this energetic penalty during the subsequent cycle. Simultaneously, the string switches the self-switching shim at $\mathrm{A}_2\mathrm{B}_2$ to the M-state.

The process repeats during the third compression cycle: The previously switched shims at the diagonal hinges suppress bending at the first two vertices, advancing the fold to hinges $\mathrm{A}_3\mathrm{B}_3$ and $\mathrm{A}_3\mathrm{B}_3'$. This switches the shims at $\mathrm{A}_3\mathrm{B}_3$ and $\mathrm{A}_2\mathrm{A}_3$ to the M-state (Fig.~\ref{fig6}f(3)).   After this cycle, all self-switching shims are in the M-state, and subsequent compression causes only the horizontal hinges to fold, producing a terminal configuration that persists until the shims are manually reset (Fig.~\ref{fig6}f(4)). The shim states therefore record the number of previous compression cycles and determine the deformation during the next cycle, enabling the sheet to function as a mechanical counter whose maximum count equals the number of $\mathrm{A}_i\mathrm{B}_i$ hinges in the array. Other arrangements of coupled buckled shims can produce similar counting mechanisms; Section~S5.4 presents an alternative design inspired by Kwakernaak et al.~\cite{kwakernaak_counting_2023}.

\section{Discussion and conclusion}
In summary, we have introduced a strategy for programming the folding response of origami sheets by embedding bistable buckled shims across their hinges. The two stable states of each shim bias the corresponding hinge toward either a mountain or a valley folding direction. Switching the state of selected shims reverses these local biases. This reprograms the global folding response and enables a single origami sheet to access multiple folding patterns.

The mechanical behavior of these reprogrammable sheets can be accurately captured using finite-element and spring-based models. In this work, we use these models to characterize how the buckled shims affect the response of individual hinges and to evaluate shim arrangements identified through physical intuition. In the future, these models could be combined with inverse-design methods to automate the search for configurations that encode multiple functionalities within a single sheet \cite{liu2026basin}. Such an approach could program not only targeted shape transformations but also changes in mechanical properties, as controlling the degrees of freedom of origami sheets has been shown to tune their rigidity \cite{chen_rigidity_2019}.

We further demonstrated two strategies for switching the shims when the hinges rotate, allowing the folding response to be passively reprogrammed through both global mechanical inputs applied at the boundaries and local inputs applied directly to the shims. By incorporating these switching strategies into origami sheets, we encode prescribed transitions between configurations, providing a foundation for new deployable and reconfigurable systems. More broadly, our approach offers a means to embed both shapes and the pathways between them into the mechanical structure of the sheet, expanding the design space of multistable origami.

\renewcommand*{\bibfont}{\fontsize{10}{10.8}\selectfont}
\bibliography{MultistableOrigami}

\section*{Methods}
\textbf{Fabrication.} Samples were fabricated by laminating compressed spring steel shims between 3D-printed PLA panels with a continuous layer of double-sided tape. A detailed description of the fabrication process is available in Supplementary information section S1.2. Specific design details of samples are shared in Supplementary information section S2-5.\\

\noindent\textbf{Mechanical testing.} Single hinges were characterized using a rheometer. Additional information about testing is available in Supplementary information section S1.2. \\

\noindent\textbf{Simulations.} We used finite-element (FE) simulations of a one-dimensional elastic beam conducted within the commercial code Abaqus/Standard to predict the torque--rotation response of a single origami hinge with embedded buckled shims. This model is described further in  Supplementary information section S1.3.\\

\noindent Furthermore a spring-based model was conceived to capture the shape of larger sheets. This is described in Supplementary information section S2.2 and the work of Liu et al. \cite{liu2026basin}.\\

\noindent Code is available from the corresponding author upon request.

\section*{Data availability}
The data that support the findings of this study are present in the paper and/or in the Supplementary Information. Additional data related to the paper are available from the corresponding author upon request. 

\section*{Acknowledgments}

We gratefully acknowledge support the ARO MURI program (W911NF-22-1-0219) and the National Science Foundation Graduate Research Fellowship Program
under Grant No. DGE 2140743 (to L.L.). Any opinions, findings, and conclusions
or recommendations expressed in this material are those of the author(s) and do not necessarily reflect the views of the National Science
Foundation. 

\section*{Aauthor contribution}
L.M.K. and K.B. conceived the research; L.M.K. designed, fabricated and conducted experimental testing; L.M.K. and D.R. performed FEA simulations; L.L. performed numerical simulations; L.M.K. and O.M. developed target angle optimization code; L.M.K and E.M. fabricated and tested self-switching origami units; K.B. supervised the work; and L.M.K., L.L. and K.B. wrote the paper with input from L.M. and D.F.\\

\noindent\textbf{Competing Interests.} The authors declare no competing interest.

\end{document}